\documentclass[twocolumn]{aastex631}

\shortauthors{Lewin et al.}
\graphicspath{{./}{/}}

\begin{document}

\title{X-ray/UVOIR Frequency-resolved Time Lag Analysis of Mrk 335 Reveals Accretion Disk Reprocessing}
\correspondingauthor{Collin Lewin}
\email{clewin@mit.edu}

\author[0000-0002-8671-1190]{Collin Lewin}
\affiliation{MIT Kavli Institute for Astrophysics and Space Research, MIT, 77 Massachusetts Avenue, Cambridge, MA 02139, USA}

\author[0000-0003-0172-0854]{Erin Kara}
\affiliation{MIT Kavli Institute for Astrophysics and Space Research, MIT, 77 Massachusetts Avenue, Cambridge, MA 02139, USA}

\author[0000-0002-8294-9281]{Edward M. Cackett}
\affiliation{Wayne State University, Department of Physics and Astronomy, 666 W Hancock St, Detroit, MI 48201, USA}

\author[0000-0002-4794-5998]{Dan Wilkins}
\affiliation{Kavli Institute for Particle Astrophysics and Cosmology, Stanford University, 452 Lomita Mall, Stanford, CA 94305, USA}

\author{Christos Panagiotou}
\affiliation{MIT Kavli Institute for Astrophysics and Space Research, MIT, 77 Massachusetts Avenue, Cambridge, MA 02139, USA}

\author[0000-0003-3828-2448]{Javier A. Garc\'ia}
\affiliation{Cahill Center for\ Astronomy and Astrophysics, California Institute of Technology, 1200 California Boulevard, Pasadena, CA 91125, USA}

\author[0000-0001-9092-8619]{Jonathan Gelbord}
\affiliation{Spectral Sciences Inc., 4 Fourth Ave., Burlington, MA 01803, USA}

\begin{abstract}
UV and optical continuum reverberation mapping is powerful for probing the accretion disk and inner broad-line region. However, recent reverberation mapping campaigns in the X-ray, UV, and optical have found lags consistently longer than those expected from the standard disk reprocessing picture. The largest discrepancy to-date was recently reported in Mrk~335, where UV/optical lags are up to 12 times longer than expected. Here, we perform a frequency-resolved time lag analysis of Mrk~335, using Gaussian processes to account for irregular sampling. For the first time, we compare the Fourier frequency-resolved lags directly to those computed using the popular Interpolated Cross-Correlation Function (ICCF) method applied to both the original and detrended light curves. We show that the anticipated disk reverberation lags are recovered by the Fourier lags when zeroing in on the short-timescale variability. This suggests that a separate variability component is present on long timescales. If this separate component is modeled as reverberation from another region beyond the accretion disk, we constrain a size-scale of roughly 15 light-days from the central black hole. This is consistent with the size of the broad line region inferred from H$\beta$ reverberation lags. We also find tentative evidence for a soft X-ray lag, which we propose may be due to light travel time delays between the hard X-ray corona and distant photoionized gas that dominates the soft X-ray spectrum below 2~keV. 
\end{abstract}

\section{Introduction} \label{sec:intro}
Understanding how material accretes onto supermassive black holes is causally linked to our grasp of feedback from active galactic nuclei (AGN) and its impact on galactic evolution as a whole. In practice, however, we are unable to spatially resolve the accretion disk in AGN, out to the broad-line region (BLR) located near the outskirts of the disk. Reverberation mapping allows us to overcome this limit by measuring the light-travel time between circumnuclear regions to inform us of their relative locations \citep[e.g. ][]{1982ApJ...255..419B, Peterson_2004, Bentz_2009, Fausnaugh_2016, Cackett_2018, Edelson_2019, Cackett_2020}. We refer to \citet{Cackett_2021} for a recent review. In essence, material at different radii from the black hole will dominate the emission in distinct wave bands, giving rise to variability in one wave band that lags or leads that of another due to the difference in light-travel time between regions. 

While the time lags between X-ray bands allow us to probe the innermost accretion flow \citep[X-ray reverberation mapping, see ][for review]{Zoghbi_2011, DeMarco_2013, Uttley_2014, Kara_2016}, the lags between longer wavelength bands grant ``sight" out to the outermost regions of the disk and the inner BLR. The accretion disk emits thermally, producing UV photons that are Compton up-scattered to X-ray energies by a region of high-energy electrons located close to the black hole, giving rise to the central X-ray-emitting region known as the corona \citep{1991ApJ...380L..51H}. These coronal X-rays then irradiate and are thermally reprocessed by the disk, which will then emit in the UV-optical-infrared (UVOIR) bands. The variability in the central corona is therefore expected to drive correlated variability that can be observed with a delay at longer wavelengths. Specifically, the X-rays first reach the inner, hotter parts of the disk before reaching the outer, colder parts. By assuming a temperature profile for a standard thin \citet{SS_1973} accretion disk ($T(R) \propto R^{-3/4}$), the lags are expected to increase in size with wavelength as $\tau \propto \lambda^{4/3}$ \citep{1998ApJ...500..162C, 1999MNRAS.302L..24C, Cackett_2007}. The normalization for this lag-wavelength relation depends on the mass and accretion rate of the black hole, as well as physical properties of the disk \citep{Fausnaugh_2016}.

Our catalog of lag measurements between the X-ray, UV, and optical have grown significantly due to recent, high-cadence, multi-color campaigns using the Neil Gehrels Swift Observatory \citep{2005SSRv..120..165B, 2005SSRv..120...95R} and ground-based telescopes, which have been carried out for 10 AGN including Mrk~335 \citep[e.g.][]{2014MNRAS.444.1469M, 2014ApJ...788...48S, Edelson_2015, Fausnaugh_2016, Cackett_2018, McHardy_2018, Edelson_2019, Cackett_2020, HernandezSantisteban_2020, Kara_2021, Kara_2022}. While these campaigns have found the time lags to roughly follow the expected $\tau \propto \lambda^{4/3}$ relation from disk reprocessing, the measured lags are on-average longer than expected, typically by a factor of 2--3. The largest discrepancy to-date was recently reported in Mrk~335 by \cite{Kara_2022}, where the lags are over an order of magnitude longer than expected given the mass and accretion rate of the source. Additionally, the discrepancies are consistently largest in the U band near 3500 \AA, where lags exceed the best-fit disk reprocessing model (even with the aforementioned longer-than-expected normalizations) by roughly a factor of 2 \citep[see Figure 5 in][]{Edelson_2019}. 

Spectroscopic monitoring of NGC~4593 by the Hubble Space Telescope revealed a clear discontinuity in the lags at the Balmer jump, corroborating the theory that the observed U-band lag excesses are due to the diffuse continuum of the BLR \citep{Cackett_2018}. Less than a year later,  \citet{2019NatAs...3..251C} used a bivariate reverberation model for distinguishing between emission components based on variability patterns in Mrk~279, from which they also concluded the too-long lags were due to contamination from reprocessing beyond the disk. Such contamination from the BLR is expected to increase the lags in all bands, with significant contamination at the Balmer jump \citep{Korista_2001, 2018MNRAS.481..533L, Korista_2019, Netzer_2022}. Since the BLR is located beyond the disk, the BLR continuum would affect the lags on timescales longer than those of the disk (up to tens of days).  

A timescale-dependent approach for computing the lags is thus valuable in order to separate the lags originating from the BLR versus the disk. A common approach used to isolate the variability on short timescales (where we expect to see contributions from the disk) is to compute the lags after ``detrending" the light curves, that is, subtracting the data by a low-degree polynomial or moving boxcar average to remove the variability operating on the longest timescales \citep[e.g.][]{McHardy_2018, HernandezSantisteban_2020, Pahari_2020, Vincentelli_2021}. For example, \citet{McHardy_2018} found that the observed UVOIR lags in NGC~4593 approach those expected from disk reprocessing  when detrending the light curves to filter out variability on timescales longer than 5 days. They also showed that reproducing the lags requires a response function consisting of a prompt response from the disk and a longer tail attributed to a distant reprocessor consistent with the BLR.  \citet{HernandezSantisteban_2020} similarly reported disk reprocessing lags as a result of detrending the light curves of Fairall~9, in effect isolating the variability present on roughly the same timescales as \citet{McHardy_2018}. 

The aforementioned works computed the lags using the popular Interpolated Cross-Correlation Function (ICCF) method of \cite{Peterson_1998}. An alternate approach is to compute the lags as a function of frequency\footnote{For clarity, all mentions of ``frequency" refer to \textit{Fourier/temporal frequency}---the inverse of which describes the timescale of the variability---as opposed to the frequency of light (wavelength is instead used in this case).} directly (the so-called frequency-resolved lags) using Fourier techniques. These Fourier lags at low-frequencies tend towards those produced by the ICCF approach \citep[][]{2013MNRAS.430..247W, Cackett_2022}, hence the use of light curve detrending to access correlated variability operating on higher frequencies (shorter timescales) when using the ICCF. An advantage to the frequency-resolved lags is that they enable a more-robust modeling of the transfer functions and thus, for instance, the geometry of the reprocessor. From modeling the frequency-resolved lags of NGC~5548 with disk reprocessing, \citet{Cackett_2022} required an additional model component to account for a distant reprocessor, again in agreement with the BLR, in order to reproduce the long lags at low frequencies. 

The aforementioned analyses of the continuum lags have shown the importance of examining the variability at different timescales (i.e frequencies), in order to separate distinct variability processes and spatial scales. However, direct application of Fourier techniques to compute the lags as a function of frequency requires the light curves to be evenly sampled. This criterion is satisfied more typically in X-ray observations (e.g. XMM-Newton), which have thus been pivotal for isolating reverberation signatures in AGN \citep[e.g.][]{Kara_2016} and black hole X-ray binaries \citep[e.g.][]{Jingyi_2022}. The required regular sampling, however, is not possible for longer wavelength observations, for instance due to weather constraints for ground observatories. As a result, alternate approaches have been devised to enable the use of Fourier techniques to irregularly sampled light curves, such as the maximum-likelihood approach first used in light curve analysis by \cite{Miller_2010}, and then expanded by \citet{Zoghbi_2013}. This approach consists of fitting an assumed model for the power spectra and cross-spectra, and thus the frequency-resolved lags \citep[as applied to UVOIR lags in][]{Cackett_2022}.  Others have fit the light curves with a maximum entropy method to recover the response function \citep{Vio_1994}.

In this paper, we overcome uneven sampling by modeling the observed variability in each wave band independently using Gaussian processes (GPs). GPs have been researched and applied extensively in the machine-learning (ML) community for decades, becoming particularly popular after  \cite{Neal1995BayesianLF} showed that infinitely complex Bayesian neural networks converge to Gaussian processes. Many even questioned if GPs would replace this fundamental ML architecture \citep{MacKay1998IntroductionTG}, given the more interpretable nature of GPs (e.g. the kernel/covariance function hyperparameters in a GP correspond to intuitive properties of the data, such as variability lengthscales and amplitudes). In the astrophysics community, the use of GPs has been growing in popularity for regression applications to sparse light curves of asteroids \citep{2021arXiv211112596W}, stars \citep{2009MNRAS.395.2226B, 2017ApJ...840...49C}, and AGN \citep{2014ApJ...788...33K, Wilkins_2019, Griffiths_2021, Lewin_2022}, and for generative modeling \citep[e.g. quasar spectra;][]{2022ApJ...938...17E}. GPs have been shown success in modeling AGN variability via the faithful reproduction of underlying autocorrelation functions \citep{Wilkins_2019, Griffiths_2021}. Most pertinent for this work is the efficacy of GPs in preserving phase/cross-correlation information between light curves: the recovery of time lags within a fractional error of a few percent has been shown using both simulations and real data \citep[][]{Wilkins_2019, Lewin_2022}.

Modeling the variability with GPs allows one to draw evenly sampled realizations of the light curves, including data in the gaps. The frequency-resolved lags are then computed from thousands of realizations, resulting in a final lag distribution. Unlike the maximum-likelihood approach, Gaussian process regression does not make assumptions regarding the cross-correlation between wave bands---the variability in any two bands is modeled independently of the other, and thus any significant cross-correlation is produced ``on its own" (i.e. as a result of structure in the original data). 

In this paper, we implement both popular approaches used for continuum reverberation mapping: the frequency-resolved lags computed with Fourier techniques and the ICCF, the latter computed using both the original and detrended light curves. We compute the frequency-resolved and ICCF lags of the well-known narrow-line Seyfert 1 (NLS1) Mrk~335 using the X-ray, UV, and optical light curves from the reverberation mapping campaign presented by \citet{Kara_2022}. In addition to finding the largest discrepancy of the lags to-date using the ICCF, they find the X-rays are not highly correlated with the UVOIR bands, until a flare is observed in all bands at the end of the campaign. When including the flare, the soft X-rays are measured to \textit{lag} the UV variability by over 10 days. This result is contrary to the disk reprocessing picture, where outward reprocessing occurs in response to variations of the central X-ray emitting region. Stated potential origins include mass accretion rate fluctuations propagating inwards in the flow and/or a vertical extent of the corona at the end of the campaign. We aim to further investigate these results using the frequency-resolved lags by probing the degree of lag contamination from the BLR using impulse response function models for standard disk reprocessing and a distant reprocessor from \citet{Cackett_2007, Cackett_2022}. 

\section{Observations} \label{sec:obs}

\begin{figure}[h!t!]
    \centering
    \includegraphics[width=\columnwidth]{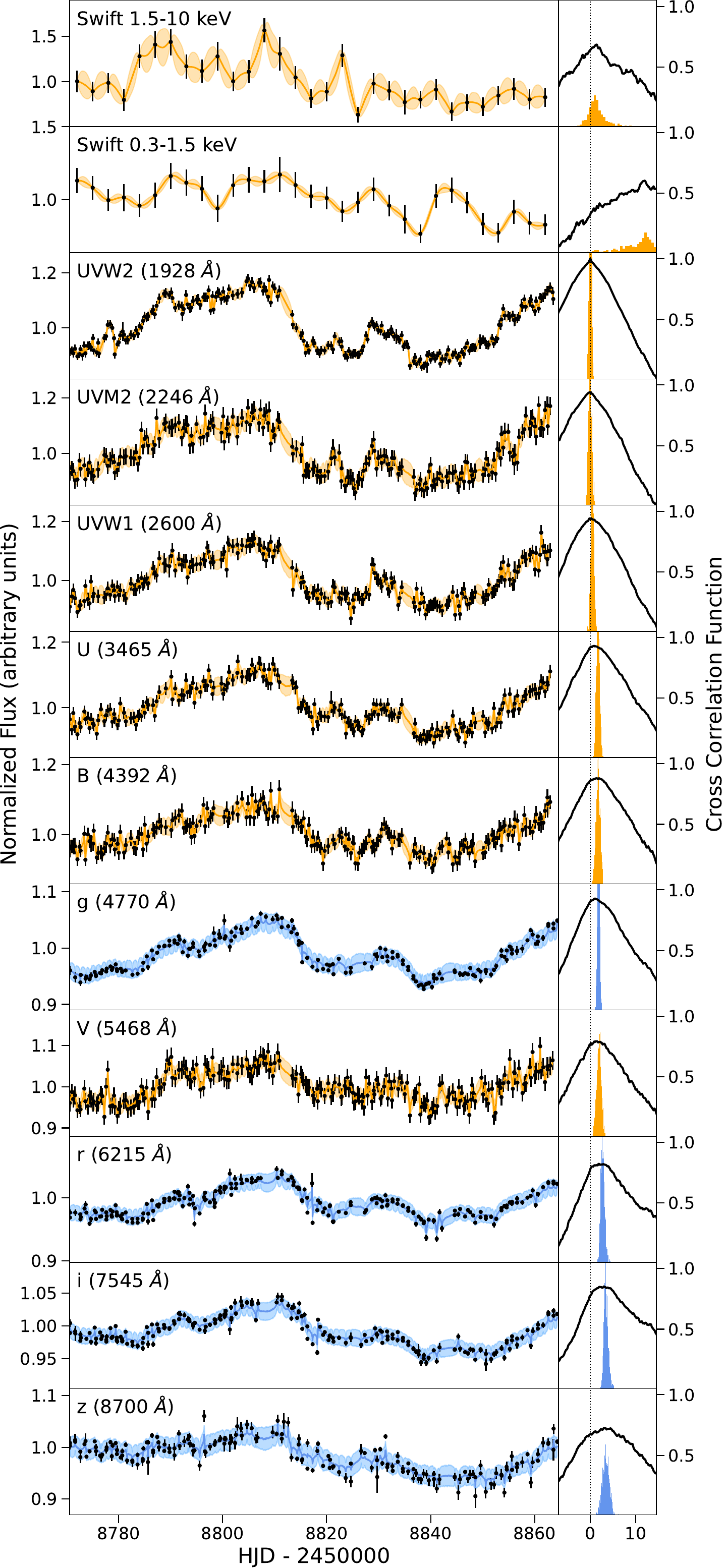}
    \caption{\textit{Left:} Light curves from the 100-day reverberation campaign presented by \citet{Kara_2022} with Swift and ground-based telescopes from MJD 2458770-2458865. The average of 1000 Gaussian process realizations is shown by solid orange and blue lines, with shaded regions indicating 1$\sigma$ in the distribution. In practice, we do not average over the realizations to compute the lags (see Section \ref{sec:fourier}). \textit{Right:} The cross-correlation function (solid black line) and the distribution of ICCF centroid lags (colored histograms), both with respect to the UVW2 band.}
    \label{fig:lcs}
\end{figure}

\cite{Kara_2022} presented the first results on a high-cadence (roughly 3 visits per day) 100-day X-ray, UV, and optical reverberation mapping campaign of Mrk~335, which began on 2019 October 14. Our analysis focuses on the data collected by Swift (XRT, UVOT) and ground-based telescopes in the 94-day time window (MJD-2450000=8770-8864) where the data are simultaneous between observatories. This window was chosen to exclude the large gap in the Swift data at the end of the campaign, as using Gaussian process regression to interpolate over the gap results in unprofitably larger uncertainties on the lags (although doing so gives consistent results). All of the light curves used in this analysis are shown in Figure~\ref{fig:lcs}, and the data are provided with the online version of this article.

The Swift X-ray light curves were produced using the Swift-XRT data products generator\footnote{ \url{https://www.swift.ac.uk/user_objects/index.php}} \citep{2007A&A...469..379E, 2009MNRAS.397.1177E}. Similar to \citet{Kara_2022}, we apply a 3-day binning of the X-ray light curves given the low count rate. We nonetheless find the CCF centroid lags and the frequency-resolved lags to agree within uncertainty from those estimated with per-observation binning. We refer the reader to \cite{Kara_2022} for details on the Swift UVOT and ground-based data reduction. This procedure includes applying a mask to mitigate the localized variations in the detector sensitivity. Out of their set of masks, we have applied the most conservative (i.e. least aggressive) mask, which filters out 4--10\% of the data depending on the band. We find that the CCF and frequency-resolved lags are very similar across mask choices, with the lags and their uncertainties deviating by less than 10\% even when using the most aggressive mask. 

The ground-based observations were carried out by the following observatories/telescopes: Las Cumbres Observatory 1m network \citep{2013PASP..125.1031B}, Liverpool Telescope 2m \citep{10.1117/12.551456}, San Pedro M\'artir Observatory 1.5m \citep{10.1117/12.926471, 10.1117/12.926927}, Wise Observatory 18-inch \citep{2008Ap&SS.314..163B}, and the Zowada Observatory 20-inch \citep{2022PASP..134d5002C}. using the SDSS \textit{g'r'i'z'} and Pan-STARRS $\textit{z}_s$ filters, with the measurements in both the SDSS and Pan-STARRS z-labeled filters combined. We henceforth refer to the filters as the \textit{griz} bands. Like \cite{Kara_2022}, we use the Swift UBV bands instead of the ground-based data collected in the SDSS \textit{u'} (poor signal-to-noise) and Johnson BV bands (poor time sampling versus Swift). 

\section{Fourier-resolved Timing using Gaussian Processes} \label{sec:fourier}

We aim to compute the frequency-resolved time lags between each wave band and a common reference band (the UVW2 band) using a Fourier-based approach in order to decompose the correlated variability occurring on different timescales/frequencies. This method requires regularly sampled data (i.e. without gaps), which is not the case for our light curves. We overcome this limitation by modeling the observed variability in each wave band independently using Gaussian processes, which allows us to then draw continuous light curve realizations, including data in the gaps, from which we compute the frequency-resolved lags.

While we refer to \citet{GPbook, Wilkins_2019, Griffiths_2021} for a more detailed introduction to Gaussian processes, we provide a condensed overview here. We have a vector of count rates $\textbf{d}$ observed at times $\textbf{t}$, which we assume to be a realization from a Gaussian process. This means that the data have been drawn from a multivariate Gaussian distribution with \textit{mean function} $m(t)=\mathbb{E}[f(t)]$, where  $f(t)$ is a function of count rates ($f(t_i)$ is the observed count rate at time $t_i$), and covariance function, henceforth referred to as the \textit{kernel function}, $k(t,t’)=\mathbb{E}[(f(t) - m(t))(f(t^\prime) - m(t^\prime))]$. 

We assume that $m=0$, given that the data is first standardized as per common practice, meaning that we subtract the mean of the light curve and then divide by the standard deviation. The kernel function describes how the data deviates from the mean function and thus models the empirical variability. One must assume the data are normally distributed as well as a functional form for the kernel function, which we discuss in the following subsections (\ref{subsec:dist}, \ref{subsec:kernel}, respectively). The choice of kernel function form has been found to impact the significance of lag recovery depending on the data sampling rate \citep{Griffiths_2021,Lewin_2022}. Each functional form has its own set of \textit{hyperparameters} $\theta$, each encoding a different aspect of the variability, such as lengthscales (timescales, in our case), amplitudes, etc. The hyperparameters are determined by finding the set of hyperparameter values that maximizes the likelihood of the model given the observed data (the marginal likelihood). In practice, it is common to minimize the negative log marginal likelihood (NLML) \citep[equation 17 in][]{Griffiths_2021}. A separate Gaussian process is trained in each wave band using only the light curve of that band, so each model is entirely self-contained and independent.

We then generate realizations with count rate data $\textbf{d}_*$ by making random draws of the conditional distribution $(\textbf{d}_*|\textbf{d})$ \citep[equation 5 in][]{Wilkins_2019}, defined by the optimized multivariate Gaussian distribution and the observed data vector \textbf{d}. We draw 1000 evenly sampled realizations, including the data in the observed gaps, in each wave band of interest and the UVW2 reference band.

We apply a standard Fourier approach to the light curve realizations in order to compute the frequency-resolved lags, a method reviewed in-detail by \citet{Uttley_2014}. In summary, the cross-spectrum is computed between each of the 1000 pairs of realizations (one realization in the band of interest, the other in the UVW2 reference band). This number of realizations was selected based on the convergence of the resulting lag distribution. The cross-spectrum is then binned into coarser frequency bins (centered at frequency $\nu$). The phase of the binned cross-spectrum is then converted to a time lag by dividing by $2\pi\nu$, resulting in a \textit{lag-frequency spectrum} between each pair of realizations. The final lag-frequency spectrum and its $1\sigma$ uncertainties are given by the mean and standard deviation of the 1000 lags in each frequency bin.

The architecture used for simulating time series in addition to model training and subsequent interpolation was created by combining and modifying the tools from \texttt{Scikit-learn}\footnote{\href{https://scikit-learn.org/}{https://scikit-learn.org/}} and the X-ray timing analysis package \texttt{pyLag}\footnote{\href{http://github.com/wilkinsdr/pylag}{http://github.com/wilkinsdr/pylag}} \citep{Wilkins_2019}. 

\subsection{Assuming a normal flux distribution}\label{subsec:dist}

We assess the applicability of Gaussian processes for modeling our light curves by testing if the empirical flux distribution is normally distributed. In actuality, accreting black holes have been found to instead follow \textit{log-normal} flux distributions \citep{Uttley_2005}. In this case, we train the Gaussian process on the log-transformed light curves (log-transforming data that originally follows a log-normal distribution will result in data that is normally distributed) and exponentiate the realizations drawn from the conditional posterior.

We perform Kolmogorov-Smirnov (K-S) tests \citep{Massey_1951} to assess the statistical difference between the cumulative distribution of our light curves from that of a standard normal distribution (where $\mu=0$ and $\sigma=1$, as our data is first standardized). In other words, the K-S test is performed using the null hypothesis that the observed flux sample was drawn from a normal distribution. To choose whether to first log-transform the light curves, we compare the \textit{p}-value from the K-S tests when using the raw flux values versus the log-transformed flux values. For our more-coarsely sampled X-ray light curves, we instead perform a Shapiro-Wilk (S-W) test, which is more appropriate for smaller data sets ($n<50$) than the K-S test \citep[$n\geq50$;][]{Mishra_2019}.

The K-S tests result in \textit{p}-values ranging from 0.01 to 0.47, with an average \textit{p}-value of 0.12. This means that in all bands (except X-rays, where this test is not used), we \textit{cannot} reject the null hypothesis that the flux values (whether log-transformed or not) have been drawn from a normal distribution at the 1\% confidence level. The deviation from the null hypothesis is even less significant (i.e. the flux distribution better agrees with a normal distribution) when log-transforming the data, which leads to greater \textit{p}-values by 11\% on-average and above 0.05 in all cases. The S-W test performed on the X-ray bands show similar results: the \textit{p}-values for the soft X-ray band are 0.07 (raw) and 0.19 (log-transformed), and for the hard X-ray band, 0.08 (raw) and 0.48 (log-transformed). 

The K-S test has been criticized for its sensitivity to only large-scale differences (the shape and median) between the empirical (cumulative) distribution functions (EDFs) of the data and model \citep{2006ASPC..351..127B}. As an additional check for Gaussianity in the UVOIR bands, we perform Cramer-von Mises (C-vM) tests \citep{CvMtest}, which are more sensitive to both small- and large-scale differences in the EDF \citep{2006ASPC..351..127B}. The results and the conclusions that follow are consistent with those from the K-S tests. The C-vM tests result in \textit{p}-values ranging from 0.01 to 0.39, with an average \textit{p-value} of 0.11; like the K-S tests, we cannot reject the null hypothesis that the flux values (whether log-transformed or not) have been drawn from a normal distribution at the 1\% confidence level. We similarly find that log-transforming the data leads to better agreement with a normal distribution, as shown by higher p-values than those found when using the raw data by roughly 10\% on-average. 

In summary, the K-S and C-vM tests both give results from which we can conclude the data is normally distributed: for the raw data, we cannot reject the null hypothesis at a 1\% significance level, although in most cases above the 10\% significance level. Log-transforming the data results in even higher \textit{p}-values, meaning that the data itself is more likely to follow a log-normal distribution. As such, we train the Gaussian processes on the log-transformed count rates in all bands.

\subsection{Selecting the kernel function}\label{subsec:kernel}

We consider the same three common forms for the kernel function as those assessed to describe AGN variability by \citet{Wilkins_2019, Griffiths_2021, Lewin_2022}: the squared exponential (SE), rational quadratic (RQ), and Mat\'ern kernels. We refer to \citet{Wilkins_2019} for an introduction to these functional forms. 

The probability of the observed light curve data given the model is quantified by the NLML function, which is minimized during hyperparameter optimization, as introduced in the previous subsection. We thus compare the kernel forms' efficacy in modeling the observed variability by comparing their optimized NLML values. Since the source variability, data sampling, and signal-to-noise varies across wave bands, the best-performing kernel form may also vary across bands; as such, we perform this comparison in each band. 

For all but the \textit{g} and UVW2 bands, we conclude that the RQ kernel best captures the observed variability based on having the lowest optimized NLML value (NLML averaged across bands: 126.5), with the Mat\'ern-$\frac{1}{2}$ kernel inching behind (average NLML: 153.2). For the \textit{g} and UVW2 bands, the Mat\'ern-$\frac{1}{2}$ kernel instead wins, with an NLML of 32.2 averaged across these two bands versus 74.6 in the case of RQ. The difference in kernel form for modeling only these two bands could result from the higher signal-to-noise in these two bands as shown in Figure~\ref{fig:lcs}. These results are generally consistent with previous kernel comparisons for modeling AGN variability, as \citet{Wilkins_2019, Griffiths_2021, Lewin_2022} all find the RQ and Mat\'ern-$\frac{1}{2}$ kernels to perform statistically similar. Similar to the three aforementioned works, we find the SE kernel provides the poorest description of the observed light curves.

We compare the four possible combinations of the top two performing kernel forms (RQ vs. Mat\'ern-$\frac{1}{2}$ for the bands of interest and the reference band) to compare how the close-call kernel choice affects the lags and their uncertainties. We find the impacts of this choice are much less noticeable in our case than that shown in \citet{Griffiths_2021}, likely due to our light curves having less sparse sampling. The lag uncertainties are similar across the kernel forms, within 10\% on-average in the lowest frequency bin and 5\% at higher frequencies. The sizes of the lags are even more invariant, within 5\% on-average in the lowest frequency bin and 2\% at higher frequencies. 

\subsection{Simulated lag recovery} \label{subsec:simlag}
Gaussian process regression has shown success in the accurate recovery of frequency-resolved time lags in AGN light curves, with previous work spanning a considerable range of sampling rates, data gap sizes, and signal-to-noise \citep{Wilkins_2019, Griffiths_2021, Lewin_2022}. As a check for how Gaussian process regression affects time-lag recovery for our specific observations, we perform simulations similar to those of \citet{Wilkins_2019, Lewin_2022}.

We first simulate light curves with lengths, means, and standard deviations matching our observations in each band using the method of \citet{Timmer_1995}, assuming a slope of -2 in the power spectral density (PSD), which is consistent with UV/optical PSDs studied for nearby AGN \citep{2018ApJ...857..141S, 2020MNRAS.499.1998P}. The amplitude of the Fourier transform at each frequency is set to match that of the assumed PSD, with the phase of each component drawn at random from a uniform distribution. The light curves are then convolved with a time-delayed $\delta$-function of the form $\delta(t-T)$ to shift the light curves by $T=2$~days (10 days for the case of the soft X-ray band), representative of the lags measured using the CCF approach (see Section \ref{sec:results}). Instead of binning these light curves to the average sampling rates of the observations, we thinned the simulated light curves by only considering points at the same time as the observed UVW2 band and the band of interest, respectively, to best replicate the observed sampling. Noise is simulated by re-drawing each flux value from a Gaussian distribution whose mean and standard deviation is set by the original flux and uncertainty, respectively.

\begin{figure}[t!]
    \centering
    \includegraphics[width=\columnwidth]{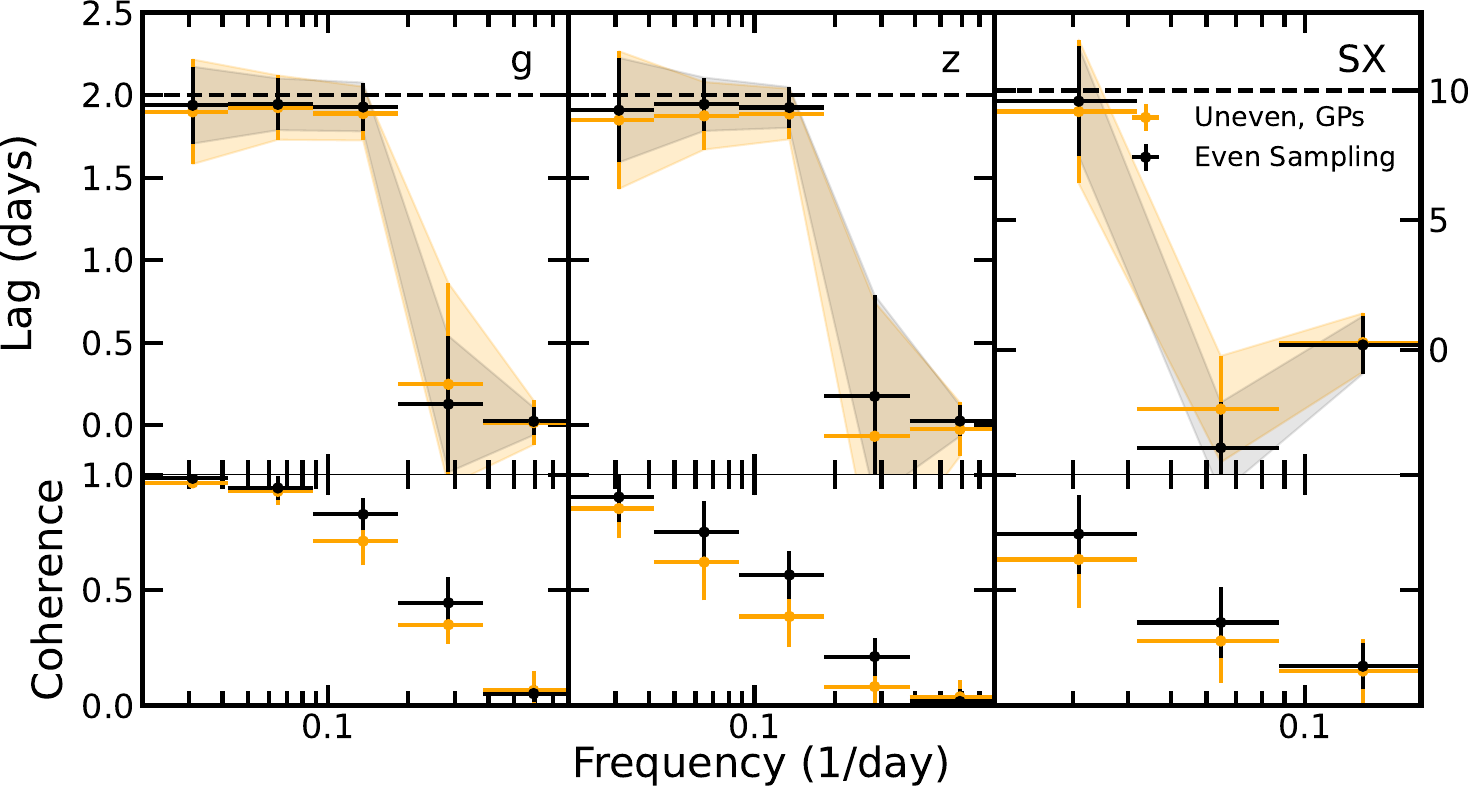}
    \caption{Distributions of lags and bias-corrected coherence from simulated light curves with a 2-day lag in the \textit{g, z} band (left, middle), and a 10-day lag in the soft X-ray band (right), with respect to the simulated UVW2 reference band. We compare the simulated lag recovery from unevenly sampled light curves, with time bins matching those of the observations and thus require the use of GPs (orange), to evenly sampled light curves (black) allowing immediate use of Fourier techniques. Shaded regions are used to visualize the overlapping lag uncertainties. The effect of applying GPs to compute the lags and their uncertainties are the most (\textit{z}, soft X-ray) and least pronounced (\textit{g}) in these bands in terms of error from the simulated lag and uncertainties compared to those computed from regularly sampled data.}
    \label{fig:sim_lfs}
\end{figure}

We then modeled the variability in each simulated light curve using Gaussian processes to compute the frequency-resolved lags, using the same frequency bins as our actual analysis. In addition to evaluating the error in the GP-recovered lags, we also compare them to the lags computed from simulated light curves that are instead uniformly binned to match the average empirical sampling rate, to which Fourier techniques can be immediately applied. This latter assessment allows us to roughly gauge the effects of GPs on the lags and their uncertainties and coherences versus those from standard Fourier techniques. The performance of these two sampling choices in recovering the simulated lags is shown in Figure~\ref{fig:sim_lfs}.

We find the impacts of Gaussian process regression on lag recovery for our observations to be generally consistent with those found by \citet{Wilkins_2019, Lewin_2022}. The lags in all bands and frequency bins using GPs agree within 5\% from those computed by immediately applying Fourier techniques to the equally sampled data. The true value of the lag lies within 1$\sigma$ from those computed with GPs in all cases (below the phase-wrapping frequency, as roughly zero-lag arises when averaging in frequency bins above this frequency). Similar to the aforementioned works, we find the use of GPs to impact the uncertainties on the lags more significantly than the sizes of the lags themselves. The uncertainties are typically larger by up to 20\% in the lowest frequency bin and larger on-average by roughly 15\%. In our worst-performing UVOIR band (\textit{z}), the uncertainties are larger by nearly 25\% on-average. 

As shown in Figure~\ref{fig:sim_lfs}, we find the use of GPs to considerably lower the measured coherence, which is expected given the increased uncertainties on the GP-recovered lags. The coherence is least affected in the lowest frequency bin, where the coherence is typically within 0.05 from those computed using the evenly sampled light curve simulations. The effects are more noticeably at higher frequencies, where the coherence is commonly 0.1-0.2 lower on-average, but still consistent within error. These results are unsurprising: the GP recovers the overall light curve shape and thus the correlated variability on long timescales. On the other hand, the shorter timescale variability in the data gaps is uncorrelated between realizations in different wave bands (as the variability amplitude drops to levels equivalent to the uncertainty), as a result of each light curve being interpolated independently without a prior dictating if one wave band should lag or lead another.

\begin{figure}[h!t!]
    \centering
    \includegraphics[width=0.6\columnwidth]{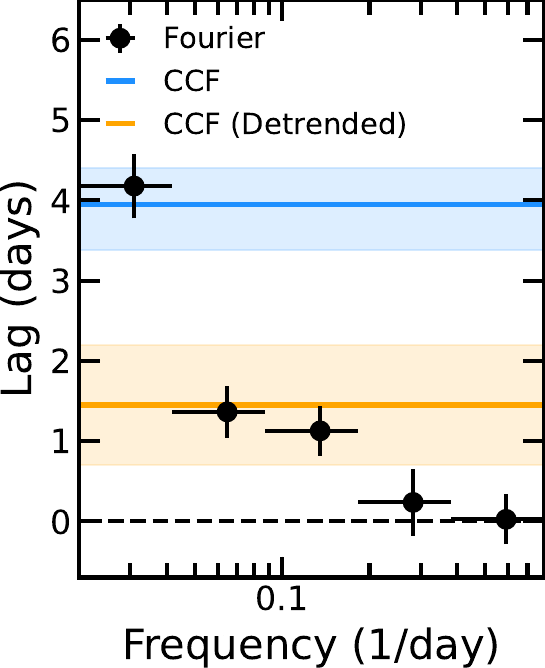}
    \caption{An exemplar band (the \textit{i} band) for comparing the lags computed with respect to the UVW2 band using Fourier techniques (black) and the (non-frequency-resolved) CCF using the original light curves (blue) and the detrended light curves (orange), with 1$\sigma$ uncertainties shown. The non-detrended CCF lag is consistent with the low-frequency component of the Fourier lags, whereas the detrended CCF lag is consistent with those at higher frequencies. See Figure~\ref{fig:lfs} for the lags computed in all bands.}
    \label{fig:lfs_i}
\end{figure}
\section{Results} \label{sec:results}
\subsection{Frequency-resolved (Fourier) time lags}
\begin{figure*}[ht!]
    \centering
    \includegraphics[width=\textwidth]{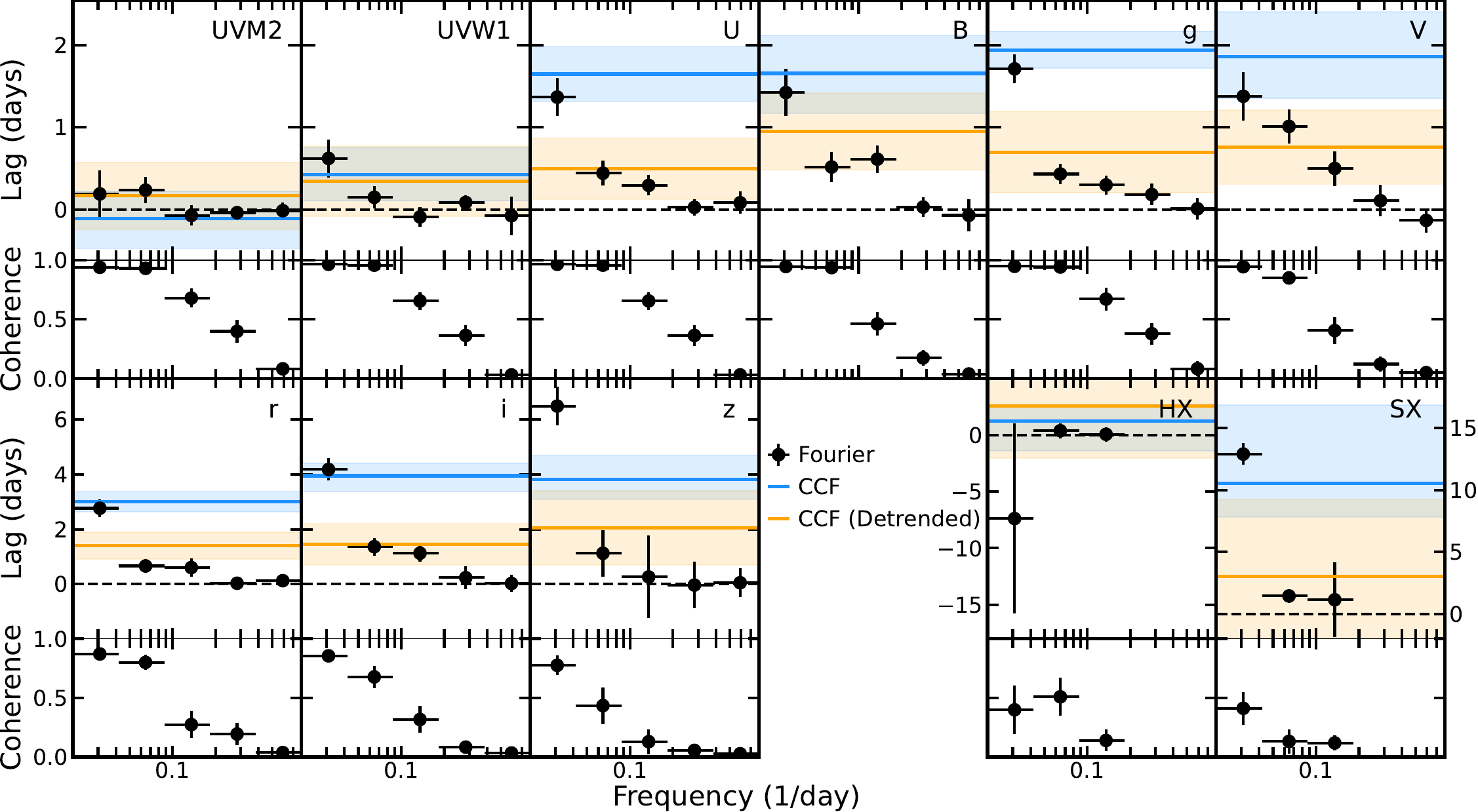}
    \caption{Frequency-resolved lags computed in each band with respect to the UVW2 reference band, with corresponding bias-corrected coherence values below.  The (non-frequency-resolved) CCF lags computed from the original light curves are shown with 1$\sigma$ uncertainties in blue, and those computed from the detrended light curves in orange. The 3-day binning of the X-ray light curves limits us to frequencies below $\sim0.2$~day$^{-1}$ (the Nyquist frequency).}
    \label{fig:lfs}
\end{figure*}
We present the frequency-resolved\footnote{All mentions of ``frequency" refer to \textit{Fourier/temporal} frequency, as opposed to the frequency of light (wavelength is instead used in this case)} time lags of Mrk~335 in the X-ray, UV, and optical, which we computed by applying Fourier techniques to Gaussian process realizations, as described in the previous section. The lags and bias-corrected coherence with respect to the UVW2 band were computed in five logarithmically spaced bins ranging from 0.02--0.8 day$^{-1}$ (except the X-ray bands, whose 3-day binning limits us to frequencies below the Nyquist frequency at $\sim0.2$~day$^{-1}$), with 1$\sigma$ uncertainties determined from the standard deviation of the lags and coherence across the 1000 GP realizations. A representative lag-frequency spectrum (the \textit{i} band) is shown in Figure~\ref{fig:lfs_i}, and the lag-frequency spectra with coherences for all wave bands are shown in Figure~\ref{fig:lfs}. The shape of the lags as a function of frequency resembles that of the lags measured in NGC~5548 by \citet{Cackett_2022} in that the size of the lags decreases with frequency. 
\begin{figure*}[t!]
    \centering
    \includegraphics[width=0.8\textwidth]{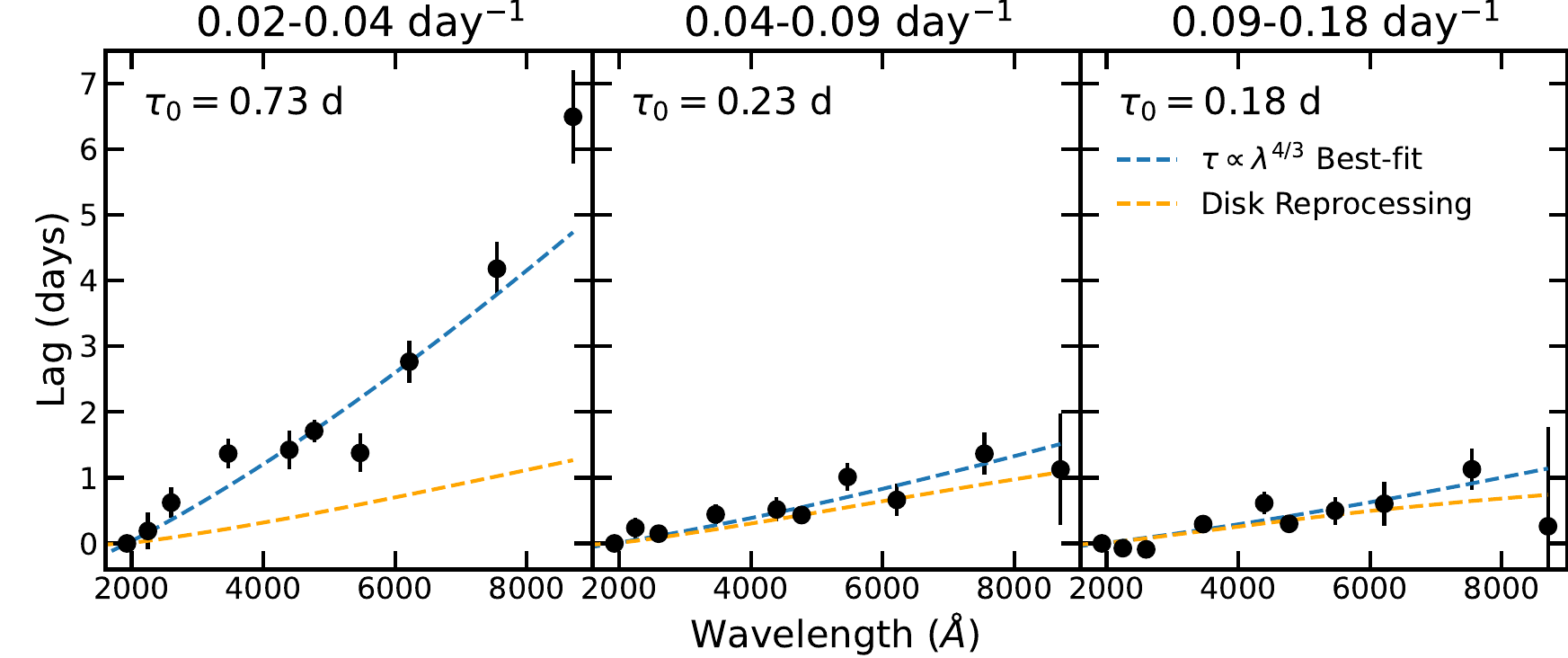}
    \caption{The frequency-resolved lags in the lowest three frequency bins, fit with a $\tau \propto \lambda^{4/3}$ relation with best-fit normalization values ($\tau_0$) shown (dashed blue line). The dashed orange line shows the expected lag-wavelength relation for this source using the disk reprocessing model in each frequency range. At higher frequencies, the lags rapidly approach the expected lag-wavelength relation for this source, with the lags in the $0.09-0.18$~day$^{-1}$ range roughly consistent with the disk reprocesing model.  If this discrepancy due to contamination from a distant reprocessor, then it is occurring on timescales longer than $1/0.09 = 11.5$~days, consistent with the radius of the BLR based on the 13.9-day H$\beta$ lag \citep{Grier_2012}. }
    \label{fig:lws}
\end{figure*}
The coherence also decreases with frequency, although the coherence in the two lowest-frequency bins is typically very high ($>0.9$). The coherence is slightly lower in the longer optical bands (\textit{riz}; $<0.85$) and substantially lower ($<\sim0.5$) in the X-ray bands. We show in Section \ref{subsec:simlag} that simulated lags can indeed be recovered in these cases, despite the lower coherence due to coarser data sampling, signal-to-noise, and the use of Gaussian processes in these bands. The low-frequency coherence measured from the actual \textit{riz}- and soft X-ray data are within error (or just nearly, within $1.1\sigma$ in the case of soft X-rays) from the simulated values after the use of GPs. It is thus difficult to conclude whether the lower coherence values that we measure are intrinsic (due to incoherent processes), given that simulating the data sampling/quality plus the use of GPs can nearly reproduce the observed coherence. Nonetheless, the lags are successfully recovered within $1\sigma$ in these cases, regardless of the low coherence exacerbated by applying GPs.

Figure~\ref{fig:lws} presents the lags in the lowest three frequency bins as a function of wavelength.  All sets of UVOIR lags increase with wavelength and roughly follow the expected $\tau \propto \lambda^{4/3}$ relation for reprocessing by a standard Shakura and Sunyaev disk \citep{Cackett_2007}. We independently fit each set of lags as a function of wavelength with the function $\tau = \tau_0 [(\lambda/\lambda_0)^{4/3} - 1]$, where $\lambda_0=1869 \text{\AA}$ is the rest-frame wavelength of the UVW2 reference band and $\tau_0$ is the normalization fit to the data. For the frequency-resolved lags, we compare the lags in each frequency range to those expected from disk reprocessing by modeling the interband impulse response functions for this source, as described in Section \ref{sec:modeling}. 

The lowest-frequency lags show the largest departure from those expected from disk reprocessing: the observed lags at this frequency are a factor of 3--7 longer ($\sim$4.5 on average) than expected. At higher frequencies, the normalization decreases to rapidly approach the expected lag-wavelength relation for this source, with the lags in the $0.09-0.18$~day$^{-1}$ range (right-most panel of Figure~\ref{fig:lws}) roughly consistent with the lags expected from disk reprocessing. 

\citet{Kara_2022} found the CCF lags in the U band ($3465 \text{\AA}$) to exceed the disk reprocessing model ($\tau \propto \lambda^{4/3}$) in this source by $\sim$30\%. Lag excesses in this band have been observed in nearly all sources observed in reverberation campaigns using Swift and ground-based monitoring \citep{Cackett_2018, Edelson_2019, HernandezSantisteban_2020, Vincentelli_2021}. We observe a similar U-band excess in the lowest frequency bin with high coherence (0.95), where the U-band lag is nearly 60\% longer than the $\lambda^{4/3}$ best-fit. Similar to the overall discrepancy from disk reprocessing, the U-band excess is resolved at higher frequencies (again, the $0.09-0.18$~day$^{-1}$ range), where it lies within $1\sigma$ from the $\lambda^{4/3}$ best-fit.

\citet{Kara_2022} also found the soft X-ray (0.3--1.5 keV) variability to \textit{lag} that observed in the UVW2 band, a result in-tension with the standard reprocessing model. We find agreeing evidence for this in the lowest frequency bin, where the soft X-ray band lags the UVW2 by $\sim13$~days. If we recompute the lags without the flare (MJD-2450000=8770-8850), the soft X-rays still lag the UVW2 by roughly 11 days, but with even higher coherence (0.63). This result is discussed further in Section \ref{sec:discuss}. The hard X-rays (1.5--10 keV), however, are not lagging the UVW2, and are actually more likely to be leading the UVW2 reference band as expected, although both the frequency-resolved and CCF lags in the hard band are consistent with zero lag. 

\subsection{Non-frequency-resolved (CCF) time lags} \label{subsec:ccf}
\begin{figure*}[ht!]
    \centering
    \includegraphics[width=\textwidth]{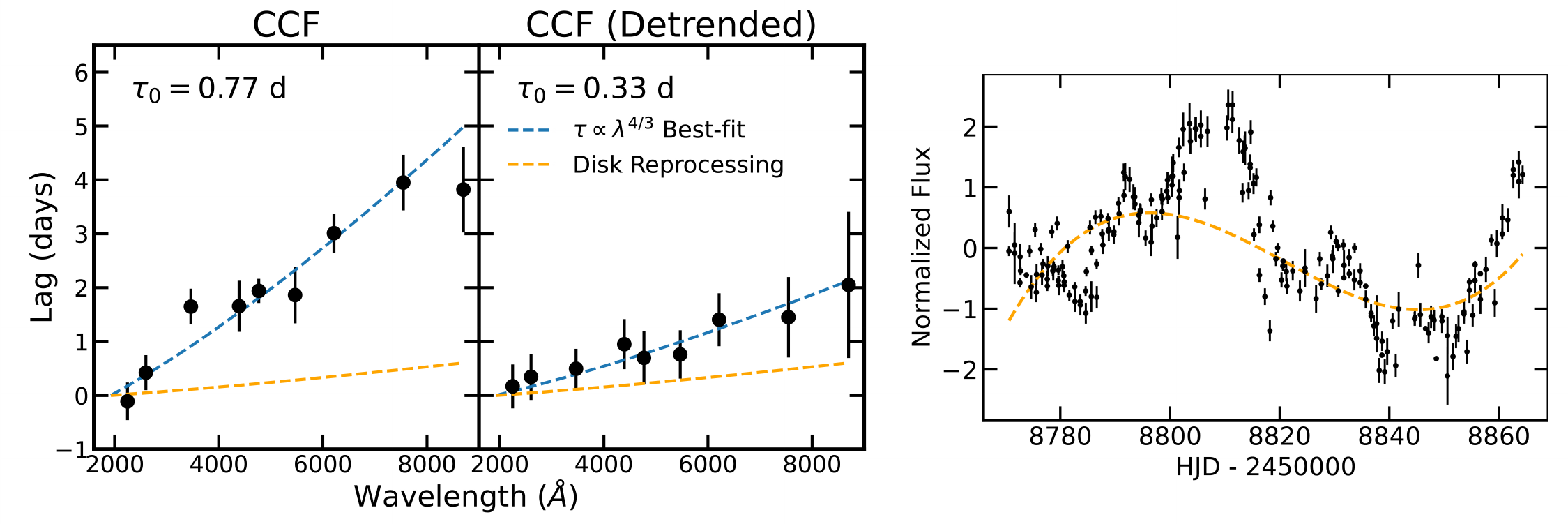}
    \caption{The CCF lags computed using the original light curves (left) and the detrended light curves (middle), both fit with $\tau \propto \lambda^{4/3}$ relations with normalization values ($\tau_0$) displayed (blue). For comparison, the lag-wavelength relation expected from disk reprocessing is shown (orange), whose normalization is $\tau_0=0.09$~days. To illustrate detrending, the \textit{i} band light curve (right, black) and the best-fit cubic polynomial that the data will be subtracted by (orange) are shown.}
    \label{fig:lws_detrend}
\end{figure*}
We also calculated the time lags between each band and the UVW2 reference band using the Interpolated Cross-Correlation Function (ICCF) method of \citet{Peterson_1998}. The ICCF is computed by shifting one of the light curves and determining the correlation coefficient by linearly interpolating the other light curve. Uncertainties on the lags were estimated using a Monte Carlo method using the flux randomization/random subset selection (FR/RSS). This method consists of generating 1000 light curve realizations with a random subset of the original data points. Each flux measurement is then redrawn from a normal distribution whose mean and standard deviation are set to the observed flux and associated uncertainty. The CCF and its centroid lag value is computed for each realization, resulting in a distribution whose median and 16\% and 84\% quantiles gives the final lags and their uncertainties. The entirety of this procedure was carried out using PyCCF\footnote{PyCCF: \url{http://ascl.net/code/v/1868}} \citep{PyCCF}. 

We apply this technique to both the original light curves and the ``detrended" light curves, meaning that each light curve is independently fit and then subtracted by a low-degree polynomial, in our case a cubic polynomial (of the form $at^3+bt^2+ct+d$ for time $t$ and parameters $a,b,c,d$). Detrending the light curves effectively removes the variability operating on the longest timescale (i.e. lowest frequency) in each light curve \citep[see][]{1999PASP..111.1347W}. The chosen degree of polynomial best reproduces the overall shape of the light curves based on mean-squared error. An illustrative example of a best-fit polynomial to the data is shown in Figure~\ref{fig:lws_detrend}.  

As an additional point of comparison between methods, we also repeated the procedure above, instead computing the lags using the Discrete Correlation Function (DCF) method of \cite{1988ApJ...333..646E}, which computes the CCF between two unevenly sampled light curves by binning the CCF itself. The DCF lags, computed using the original and detrended light curves, were consistent within error from the CCF lags shown here; as such, only the CCF lags are presented.

The resulting CCFs and the distributions of ICCF centroid lags computed using the original light curves are shown in Figure~\ref{fig:lcs}.  The lags and their uncertainties for both the original and detrended light curves are shown in Figure~\ref{fig:lfs_i} (for the exemplar \textit{i} band) and Figure~\ref{fig:lfs} (for all wave bands). The (not-detrended) CCF lag results are generally in agreement with \citet{Kara_2022} as expected, given that we used roughly 85\% of the same data. We instead do not use the data after the large gap in the Swift data, since implementing Gaussian process regression over this large gap significantly increased uncertainties on the lags (but gave results consistent within error). Similar to \citet{Kara_2022}, we find the maximum correlation coefficient (commonly denoted $R_{max}$) values are generally high ($>0.8$), but slightly lower in the \textit{z} band (0.73) and lowest in the X-ray bands ($0.60-0.68$). The centroid lags are consistently longer than the ICCF peak lags (by $\sim$ 20\% on average), indicating that the transfer function is asymmetric with a tail to long lags, providing evidence for reprocessing on long timescales \citep{Cackett_2022}.

We find that the CCF lags computed using the original light curves are consistent with the Fourier lags in the lowest frequency bin, except in the \textit{z} band, as expected given that the low-frequency lags have been found to tend to the CCF centroid \citep[see e.g.][]{2013MNRAS.430..247W}. The CCF lags computed using the detrended light curves are generally consistent with the Fourier lags at higher frequencies (by a factor of 2--4 higher than those matching the not-detrended CCF). These results motivate investigation as to what processes are operating on different timescales, for instance reprocessing off of the BLR versus the disk. Such contamination from the BLR would most significantly impact these low frequencies given the observed radius of the BLR, and is thus a potential origin for the largest discrepancy from standard disk reprocessing seen to-date \citep[as first reported by][]{Kara_2022}. 

Figure~\ref{fig:lws_detrend} shows the CCF lags computed from both the original and detrended light curves as a function of wavelength. Again, the lags roughly follow the expected $\tau \propto \lambda^{4/3}$ relation for standard disk reprocessing and are thus fit with the function $\tau = \tau_0 [(\lambda/\lambda_0)^{4/3} - 1]$, where $\lambda_0=1869 \text{\AA}$ is the rest-frame wavelength of the UVW2 reference band and $\tau_0$ is the normalization fit to the data. The expected value for the normalization depends on the mass, accretion rate, and physical properties of the disk \citep[see equation 12 in][]{Fausnaugh_2016}. Assuming a black hole mass of $(1.69\pm0.17)\times 10^7 M_\odot$ \citep{Grier_2012} and $L/L_{\text{Edd}}=0.07$ \citep{Tripathi_2020}, the expected normalization of the CCF lags is $\tau_0=0.09\pm0.01$~days. 

The discrepancy from the expected normalization is most substantial in the CCF lags computed from the original light curves ($\tau_0=0.77$~days): the normalization is longer than expected by a factor of almost 9. This disagreement more than halves when computing the CCF lags using the detrended light curves, which results in a considerably smaller normalization ($\tau_0 = 0.33$~days). While the detrended lags are thus more consistent with those expected from disk reprocessing, they are still ``too-long" by a factor of 3.  

Similar to the low-frequency lags, we observe a U-band  ($3465 \text{\AA}$) excess in the (non-detrended) CCF lags, where the U-band lag is nearly 80\% longer than even the $\lambda^{4/3}$ best-fit. Just as the U-band lag excess is resolved in the Fourier lags at higher frequencies, the U-band excess is resolved when detrending the light curves in that the lag becomes within $1\sigma$ from the $\lambda^{4/3}$ best-fit. 

\section{Modeling the Frequency-resolved Lags} \label{sec:modeling}

\begin{figure*}[ht!]
    \centering
    \includegraphics[width=\textwidth]{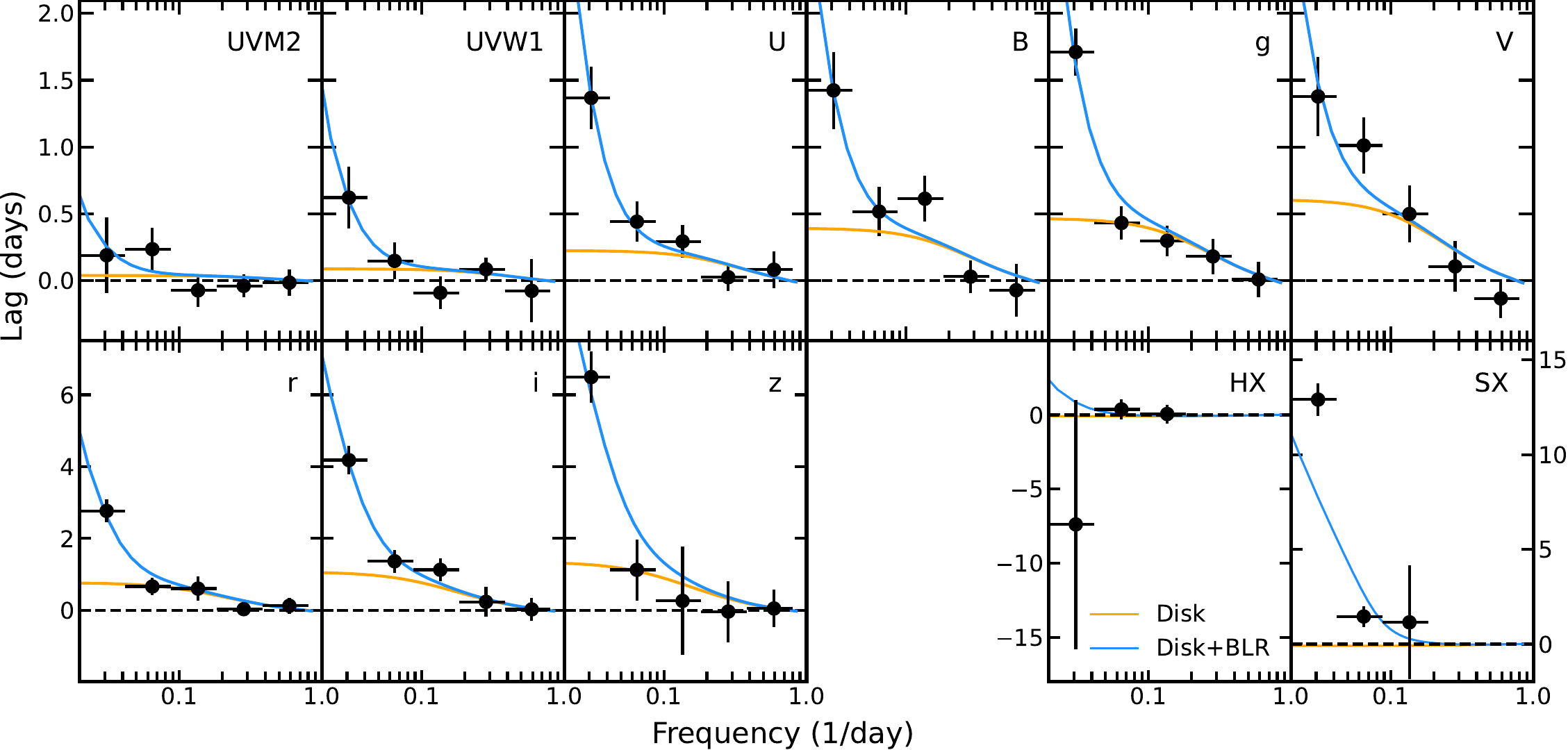}
    \caption{Frequency-resolved lags expected from the standard disk reprocessing model given the mass and accretion rate of Mrk~335 (orange), which provides a poor description of the lags at low frequencies. The fit significantly improves when including a log-normal component to account for additional contribution to the lags from a distant reprocessor located at a radius set to the measured 13.9-day H$\beta$ lag (i.e. the BLR) (blue). Fitting for the radius of the reprocessor still results in a radius consistent with the BLR, see Section \ref{sec:modeling}}
    \label{fig:lfs_modeling}
\end{figure*}

\begin{figure}[t!]
    \centering
    \includegraphics[width=\columnwidth]{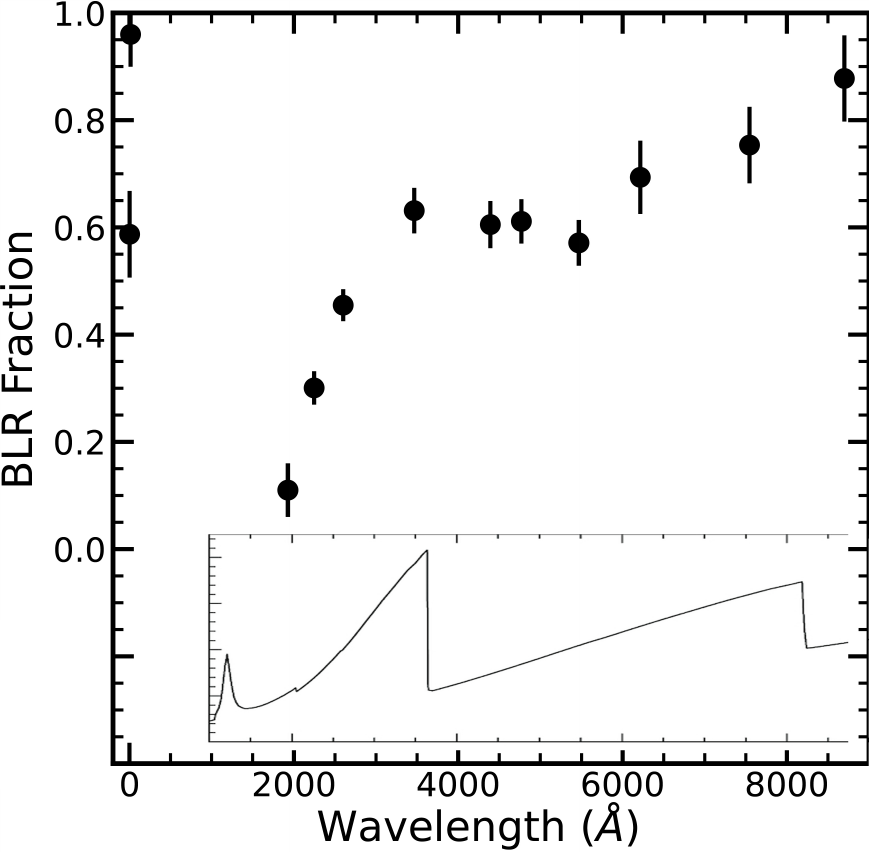}
    \caption{Fraction of the total (disk+BLR) model made up from the BLR component, denoted $f$ in Equation \ref{eq:totresp}. The BLR fraction shows a local maximum in the U band near 3500 \AA. As a point of visual comparison, the inlay at the bottom is Figure 9 from \cite{Korista_2019}, which shows the ratio of the diffuse continuum emission to the total SED as a function of wavelength, with x-axis aligned to match our plot.}
    \label{fig:blrfrac}
\end{figure}

We aim to model the Mrk~335 lag-frequency spectra presented in Figure~\ref{fig:lfs} by first modeling the expected frequency-resolved lags from standard disk reprocessing. We apply the model described in \citet{Cackett_2007}, which is characterized by disk temperatures (at an arbitrary radius of 1 light-day) in a faint and bright state ($T_B, T_F$). The CCF lags as a function of wavelength depend on these temperatures \cite[equation 13 in][]{Cackett_2007}. Similar to \cite{Cackett_2022}, we set the temperatures to match the lag-wavelength relation expected from standard disk reprocessing \citep{Fausnaugh_2016}, which we determine by computing a normalization of $\tau_0=0.09\pm0.01$~days, assuming a black hole mass of $(1.69\pm0.17)\times 10^7 M_\odot$ \citep{Grier_2012} and $L/L_{\text{Edd}}=0.07$ \citep{Tripathi_2020}. This results in temperatures $T_F=2670$ K and $T_B=4530$ K.

Similar to \cite{Cackett_2022}, we compute the frequency-resolved lags expected from the aforementioned disk reprocessing model by computing the impulse response function at each wavelength of interest \citep[equation 7 in][]{Cackett_2007}, assuming an inclination of 57$^\circ$ \citep{Wilkins_2015}. This approach assumes that the light curve in each band is reprocessed with respect to (i.e. is a convolution of) a driving light curve. In order to account for the UVW2 band also being a reprocessed light curve (i.e. not the driving light curve), we multiply the complex conjugate of the UVW2 transfer function (the Fourier transform of the impulse response function) by the transfer function of each band of interest \citep[see][for more details]{Cackett_2022}. The phase lag of this product of transfer functions (which is a cross-spectrum) is then converted to a lag-frequency spectrum by dividing by $2\pi\nu$, where $\nu$ is the frequency of each bin.

The resulting lags from this model are shown in Figure~\ref{fig:lfs_modeling}. The disk model response function has an immediate peak and tail \citep[as shown in Figure 2 of][]{Cackett_2007}, giving rise to roughly constant lags at low frequencies and lags that decrease above some frequency. Equivalently, more reprocessing is seen when moving farther out in the disk (longer lags), until reaching a radius of maximum reprocessing. Moving out beyond this radius, reprocessing becomes negligible as the impulse response approaches zero, resulting in a roughly constant lag at low frequencies. 

The disk model provides a poor description of the low-frequency lags, where the model consistently undershoots the lags, often by a factor of 3 to 4. As a result, the shape of the model differs noticeably from the lags: the disk reprocessing model flattens below roughly 0.1 day$^{-1}$, whereas the observed lags increase, often rapidly, at these lower frequencies. The disk model, however, is much better at reproducing the high-frequency lags. The reduced chi-squared is $\chi^2_\nu = 520.5/51 = 10.2$. The high $\chi^2$ value is exacerbated by the inability of the model to recreate the long soft X-ray lag, since reprocessing anticipates a negative lag with respect to the UVW2 band. Without the X-ray bands, the fit statistic improves to $\chi^2_\nu = 285.2/45 = 6.3$. These results are in agreement with \cite{Cackett_2022}, who similarly found that the low-frequency lags of NGC~5548 could not be adequately reproduced by disk reprocessing. They instead found the observed lags to be significantly better-described when including an additional impulse response function to account for potential lag contributions from a distant reprocessor (consistent with the BLR). We apply this procedure to our data, taking the final impulse response function ($\psi_{\text{tot}}$) to be a combination of that from disk reprocessing ($\psi_{\text{disk}}$) and a distant reprocessor representing the BLR ($\psi_{\text{BLR}}$):
\begin{equation} \label{eq:totresp}
    \psi_{\text{tot}}(t) = (1-f)\psi_{\text{disk}}(t)+f\psi_{\text{BLR}}(t)
\end{equation}

\noindent where $f$ is the fractional contribution of the BLR to the total impulse response function. All impulse response functions are normalized to have a total area of 1. We use the same simple model as \citet{Cackett_2022} for an extended reprocessor, a log-normal impulse response:
\begin{equation}
    \psi_{\text{BLR}}(t) = \frac{1}{S\sqrt{2\pi}t}\exp \left[ -\frac{(\ln(t)-M)^2}{2S^2} \right]
\end{equation}

A log-normal impulse response is an analytic prescription to model reprocessing by the BLR and is a smoother alternative to the top-hat response function expected from reprocessing by a spherical shell \citep{Uttley_2014}. We refer the reader to Figure 5 in \citet{Cackett_2022} how the final impulse response function varies for different BLR fraction values using the same models for the disk and BLR as this paper, albeit with slightly different $T_F, T_B$ values. 

The median of the log-normal shaped response is $e^M$, so we set $M=\ln(13.9)$ in an initial test using the 13.9-day H$\beta$ lag from previous BLR reverberation mapping campaign results \citep{Grier_2012}. The standard deviation $S$ is initially set to 1 for simplicity. Just as the contribution from the BLR diffuse continuum varies across wavelengths \citep{Korista_2001, Korista_2019}, we fit the BLR fraction of the total response function to the observed lags in each band independently. 

The resulting best-fit of the disk+BLR model is shown in Figure~\ref{fig:lfs_modeling}. This model provides a much better description of the observed low-frequency lags than the disk reprocessing model alone. Including the BLR component (with median fixed at 13.9 days) improves the fit statistic to $\chi^2_\nu = 73.0/39 = 1.9$ from $\chi^2_\nu = 520.5/51 = 10.2$ in the case of the disk reprocessing model. The resulting BLR fractions as a function of wavelength are shown in Figure~\ref{fig:blrfrac}. The BLR model component contributes more to the final model at longer wavelengths, but shows evidence for a local maximum in the U band, consistent with the U-band lag excess thought to originate from the Balmer jump in the BLR diffuse continuum. 

Even with the BLR component, the model is unable to replicate the long soft X-ray lag. While we do not expect a significant contribution to the soft X-ray band by the BLR, we use the model component as a proxy for the radius required to produce such a lag. If we re-fit excluding the X-ray bands, the reduced chi-squared improves to $\chi^2_\nu = 29.0/35 = 0.83$ from $\chi^2_\nu = 285.2/45 = 6.3$. 

We also refit the data with the combined disk+BLR model, this time fitting for the median ($e^M$) and standard deviation ($S$) of the distant reprocessor model, instead of assuming the observed 13.9-day median. The BLR fraction is again fit in each band independently. The disk component remains the same as before, with values dictated by the mass and accretion rate. We spanned values for $e^M$ from 0.5--25 days and $S$ from 0.1--5 days, with and without the X-ray bands. Fitting without the X-ray bands results in best-fit values and $1\sigma$-uncertainties of $e^M=15.4^{+2.4}_{-3.0}$~days and $S=0.9^{+0.2}_{-0.1}$~days. The radius of the extended reprocessor inferred from fitting the UVOIR lags is thus in agreement with that of the BLR based on the observed 13.9-day H$\beta$ lag \citep{Grier_2012}. The fit statistic ($\chi^2_\nu = 26/33 = 0.79$) is within 5\% of that found when using the measured $e^M=13.9$~days and assuming $S=1$ ($\chi^2_\nu = 0.83$). When including the X-ray bands, the extended reprocessor component requires a slightly larger radius in-attempt to produce the measured soft X-ray lag at $e^M=16.4^{+1.1}_{-1.2}$~days and $S=0.9\pm0.1$~days.

In order to probe the radii of reprocessing required to reproduce the lags at each wavelength independently, we fit the lags but allow the parameters of the extended reprocessor model to differ across wave bands. This results in the median of the extended reprocessor to be consistent with the observed H$\beta$ lag/BLR in all cases except the soft X-ray band, although the parameters are less tightly constrained. We are unable to model the long soft X-ray lag with a single shared $e^M$ and $S$; instead, fitting the soft X-ray band alone requires a radius larger than that of the BLR, with $e^M = 18.7^{+1.2}_{-1.4}$~days and $S=0.9\pm0.1$~days. This result is discussed further in Section \ref{sec:discuss}.

\citet{Cackett_2022} also attempted to model the frequency-resolved lags of NGC~5548 with reprocessing off of a significantly larger disk. We performed a similar final test using only the disk reprocessing model, but now with a larger disk (i.e. normalization): instead of setting the disk temperatures by fitting the lags expected given the mass and accretion rate of the source, we instead fit the temperatures to our measured CCF lags, resulting in much hotter temperatures $T_F=36800$~K, $T_B=46200$~K than before. This model provides a considerably worse description of the lags than the combined BLR+smaller-disk model, resulting in a fit statistic of $\chi^2_\nu = 510.2/51 = 10.0$ when including the X-ray bands and $\chi^2_\nu = 97.4/45 = 2.16$ when excluding the X-ray bands (versus $\chi^2_\nu = 0.83$ from the smaller disk+BLR model). Regardless, the normalization required to reproduce the CCF lags with thin-disk reprocessing ($\tau_0=0.77$~days) would require $L/L_{\text{Edd}} = 43.8$ if we assume a black hole mass of $(1.69\pm0.17)\times 10^7 M_\odot$ \citep{Grier_2012}, orders of magnitude higher than the observed value for this source and the accretion rate at which the thin-disk model is expected to hold \citep{Fausnaugh_2016, Tripathi_2020}. 

\section{Discussion} \label{sec:discuss}

A recent reverberation mapping campaign of the well-known NLS1 Mrk~335 in the X-ray, UV, and optical \citep{Kara_2022} resulted in several interesting findings. While all recent reverberation mapping campaigns have reported UVOIR continuum lags longer on-average than those expected from standard disk reprocessing, the Mrk~335 lags were found to be longer by a factor of 5--12---the largest discrepancy to date. These long lags are often thought to be the result of additional contribution to the lags from the diffuse continuum of the BLR \citep{Korista_2001, Korista_2019}, which also explains why the most significant lag excesses are consistently observed near the Balmer jump \citep{Cackett_2018}. If the BLR interpretation is correct, one would expect reprocessing from the BLR to dominate on long timescales, and that from the disk on short timescales, thus motivating a frequency-resolved approach.

In this paper, we presented and modeled the frequency-resolved lags of Mrk~335, as shown in Figures \ref{fig:lfs} and \ref{fig:lfs_modeling}, which were calculated by applying Fourier techniques to Gaussian process realizations in order to overcome uneven sampling. We compare these results to those computed using the (non-frequency-resolved) Interpolated Cross-Correlation Function (ICCF) method commonly used for reverberation mapping measurements. We compute the CCF lags from both the original and ``detrended" light curves, meaning that we fit and subtract the light curves by a low-degree (cubic) polynomial to remove the variability on the longest timescales. 

The lags are systematically longest in the lowest frequency bin, and decrease in size at higher frequencies, similar to the frequency-resolved lags in NGC~5548 \citep{Cackett_2022}. The lags in the lowest frequency bin (0.02--0.04 day$^{-1}$) are often noticeably longer, often by a factor of 3-4, than the lags observed at any higher frequency. As a result, the Mrk~335 lags show a steeper slope below 0.1 day$^{-1}$ than those observed at roughly the same frequency in NGC~5548. This could be indicative of stronger contamination by the BLR in Mrk~335, or related to differences in the intrinsic variability of these systems at low frequencies. In addition, the CCF lags computed from the original light curves in this source are almost always consistent with the lags in the lowest frequency bin (see Figure~\ref{fig:lfs}). If the continuum of the BLR is contaminating the low-frequency lags and thus the CCF lags, then this could contribute to the CCF lags in this source showing the largest discrepancy in the lags from disk reprocessing yet \citep{Kara_2022}. In either case, studying more objects with a similar approach will allow us to better differentiate between BLR contamination strength and properties of the source variability at low frequencies.

In order to probe contamination of the low-frequency lags, we first modeled the frequency-resolved lags expected from reprocessing off a standard \citet{SS_1973} accretion disk, using the impulse response function model of \citet{Cackett_2007}. We set the model temperatures to match the expected CCF lags computed using the lag normalization from \citet{Fausnaugh_2016} given Mrk~335's mass \citep[$(1.69\pm0.17)\times 10^7 M_\odot$;][]{Grier_2012} and observed $L/L_{\text{Edd}}=0.07$ \citep{Tripathi_2020}.  The lowest-frequency lags are longer by a factor of 3-7 ($\sim$4.5 on average) than those expected from disk reprocessing in this frequency range, in addition to a U-band excess characteristic of BLR contamination by roughly 80\% above even the $\tau\propto\lambda^{4/3}$ best-fit. As shown in Figure~\ref{fig:lws}, the lags rapidly approach the expected disk reprocessing lags at higher frequencies. The lags above 0.09 day$^{-1}$ are generally well-described by the disk model, including a resolution to the U-band excess. Therefore, if the discrepancy in the observed lags from disk reprocessing is due to contamination from a distant reprocessor, then it is occurring on timescales longer than $1/0.09 = 11.5$~days, which is consistent with the radius of the BLR based on the 13.9-day H$\beta$ lag observed by \citet{Grier_2012}. The CCF lags tell a similar story: detrending the light curves to remove the longest-timescale variability results in lags considerably more consistent with disk reprocessing, but the lags are still too-long by a factor of 3 (see Figure~\ref{fig:lws_detrend}).

We are unable to reproduce the measured frequency-resolved lags with standard disk reprocessing (Fig. \ref{fig:lfs_modeling}). Fitting the disk temperatures to the observed CCF lags instead of assuming values based on the mass and $L/L_{\text{Edd}}$ requires an accretion rate orders of magnitude higher than both the observed value \citep[$L/L_{\text{Edd}} = 43.8$ vs. $0.07$;][]{Tripathi_2020} and the assumptions of the thin-disk model \citep{Fausnaugh_2016}. As a result, we include the log-normal impulse response function used by \citet{Cackett_2022} to model additional lag contributions from a distant reprocessor. We first set the median of the component to match the observed 13.9-day H$\beta$ lag \citep{Grier_2012} and allow for the fractional contribution of the BLR model to the total disk+BLR model to vary across bands, as expected from the BLR diffuse continuum \citep{Korista_2001, Korista_2019}. This model provides a much better description of the UVOIR lags, especially at low frequencies ($\chi^2_\nu = 0.83$ vs. $\chi^2_\nu = 6.3$ from the disk model alone, without the X-ray bands), but is still unable to fully reproduce the long soft X-ray lag. When we fit for the radius of the distant reprocessor model, the median is constrained at $15.4^{+2.4}_{-3.0}$~days, indicating that the distant reprocessor is constrained at a radius consistent with that of the BLR.

The lowest-frequency lags increase much faster as a function of wavelength than predicted by disk reprocessing, as shown in Figure~\ref{fig:lws}. To account for this, the BLR model contributes more to the total model at longer wavelengths, as was the case for \citet{Cackett_2022} when modeling the frequency-resolved lags in NGC~5548. We report BLR fraction values similar to those reported for NGC~5548, in addition to finding a local maximum in BLR fraction in the U band as expected if BLR contamination is the culprit to the U-band lag excess. The BLR fraction increases rapidly leading up to the U band, followed by a slower rise at longer wavelengths. This shape resembles the expected contribution from the BLR diffuse continuum and its associated lags \citet{Korista_2001, Korista_2019}, although one might expect a larger drop in the B-band BLR fraction immediately following the Balmer jump. 

We emphasize that our modeling of the lags is limited in that we have applied a single model for disk reprocessing and an analytic treatment to account for contributions to the lags by an extended reprocessor. While we do explore multiple sets of parameters for the disk reprocessing model (namely, the disk temperatures expected for a black hole of this mass and accretion rate, in addition to those found from fitting the measured CCF lags), additional modeling using more physical models is warranted. This includes more complex models for the disk, such as those that include general relativity \citep{2021ApJ...907...20K, 2021MNRAS.503.4163K}, and more physical models for the BLR \citep{Korista_2019, 2020MNRAS.494.1611N}.

\subsection{X-ray variability lagging the UV}
\citet{Kara_2022} also found a low correlation between the X-ray and UVOIR bands, until a flare is observed at the end of the campaign. The soft X-rays are measured to then \textit{lag} the UV variability by over 10 days. This result is contrary to reprocessing occurring in response to variations of the central X-ray emitting region. They propose mass accretion rate fluctuations propagating inwards in the flow and/or a vertical extent of the corona at the end of the campaign as potential solutions. In our frequency-resolved lags, we also found that the soft X-ray band (0.3-1.5~keV) lags the UVW2 band by roughly 13 days. We show in Section \ref{subsec:simlag} the successful recovery of simulated lags in the X-ray bands, despite the lower coherence introduced by coarser data sampling, signal-to-noise, and the use of Gaussian processes. If we remove the flare at the end of the campaign (MJD-2450000=8770-8850), we still find a $\sim$11-day lag of the soft X-rays behind the UV with a higher coherence than before (0.62 vs. 0.5). Fitting for the radius of the distant reprocessor using only the soft X-rays results in a slightly larger radius ($18.7^{+1.2}_{-1.4}$~days) than when fitting the UVOIR bands ($15.4^{+2.4}_{-3.0}$~days), albeit within 1.5$\sigma$. 

High-resolution X-ray spectra of Mrk~335 revealed soft X-ray lines indicative of hot photoionized gas located at a radius of $\sim$7-80 light days  \citep{Longinotti_2008, 2019MNRAS.490..683P, 2021MNRAS.506.5190L}. The primary X-ray continuum will first reach the accretion disk (causing a response of the UV/optical bands), and only later reaches the more distant circumnuclear gas, responsible for the soft X-ray lines. In other words, both the accretion disk and the distant circumnuclear material are responding to variations in the primary continuum, but because the accretion disk is closer to the central source, we see the UV/optical lead the soft X-ray band. This adds complexity to the central reprocessing picture, where the X-ray variability of the corona drives (and should thus lead) the longer-wavelength variability. The location of the gas beyond the BLR would explain why the lag is seen at low frequencies, and thus why the distant reprocessor component requires a larger radius in this band only.

Unlike the soft X-ray band, the hard X-ray band (1.5-10~keV) does not lag the UV. Using XMM-Newton and NuSTAR observations of Mrk 335 taken in a similar low X-ray state, \cite{2019MNRAS.490..683P} found that the reprocessed emission from this hot, photoionized gas dominates the X-ray spectrum below $\sim$2~keV, whereas the hard X-rays represent more closely the intrinsic continuum, and therefore show little lag with respect to the UV. This could explain why only the soft X-rays are seen to lag the UV, whereas the hard X-rays are more likely to instead lead the UV. In this scenario, we would expect the soft X-rays from the distant plasma to also lag the hard X-rays from the corona. As a check, we compute the lags between the hard and soft X-ray bands using the method outlined in Section \ref{subsec:ccf}, and find (tentatively, given the data quality in both of these bands) that the soft band lags the hard by $6.8^{+12.1}_{-3.1}$~days, with a maximum correlation coefficient ($R_{max}$) of 0.61. 

We note that the soft X-ray variability lagging the UV (and tentatively the hard X-rays) due to reprocessing by hot, photoionized gas is only one possible scenario. For instance, \cite{2016A&A...596A..79S} showed that the non-zero photoionization and recombination timescale of warm absorbers could lead to a soft X-ray lag. However, given the inferred densities of typical warm absorbers, these lags are expected to be much shorter than the soft lag seen here. Another possibility is that the X-ray lag is produced from fluctuations in the mass accretion rate that flow inwards through the disk on the viscous timescale \citep[e.g.][]{1997MNRAS.292..679L, 2008MNRAS.389.1479A}, as proposed by \cite{Kara_2022}.

\section{Conclusions} \label{sec:conclusions}
We have computed the frequency-resolved X-ray and UVOIR lags of Mrk~335, which we attempted to reproduce by modeling the lags produced from reprocessing by a standard \citet{SS_1973} accretion disk. We directly compare these frequency-resolved lags to those computed using the popular ICCF method applied to both the original and detrended light curves. Here are our main results:

\begin{enumerate}
    \item We modeled the observed variability in each wave band with Gaussian processes, allowing us to generate evenly sampled realizations from which we compute the frequency-resolved lags presented in Figure~\ref{fig:lfs}. 

    \item The lowest-frequency ($0.02-0.04$~day$^{-1}$) lags are longer by a factor of 3--7 ($\sim$4.5 on average) than those expected from standard disk reprocessing, including a U-band ($\sim$3500~\AA) excess of roughly 60\% near the Balmer jump. 
    
    \item We computed the theoretical frequency-resolved time lags expected from a \citet{SS_1973} disk. We find that the high-frequency lags are well described by the disk reprocessing model, including a resolution of the U-band excess, but the low-frequency lags require an additional component (see Figure~\ref{fig:lws}).

    \item We are unable to reproduce the observed lags, especially at low frequencies, with thin-disk reprocessing models. Modeling the CCF lags with only thin-disk reprocessing requires an accretion rate orders of magnitude higher than the observed value. The CCF lags become more consistent (but not fully) with disk reprocessing after detrending the light curves, including a resolution of the U-band excess (see Figure~\ref{fig:lws_detrend}).

    \item The frequency-resolved lags are well-described when including a model component that accounts for additional contribution to lags from a distant reprocessor at a radius set to that of the broad-line region, based on previously measurements of the H$\beta$ lag (see Figure~\ref{fig:lfs_modeling}). Fitting the UVOIR lags for the radius of this component results in a value consistent with the measured 13.9-day H$\beta$ lag ($15.4^{+2.4}_{-3.0}$~days).
    
    \item The soft X-ray band (0.3-1.5~keV) lags the UVW2 band by roughly 13 days, contrary to the standard reprocessing picture. We show simulated lags are successfully recovered in the X-ray bands, despite the lower coherence introduced by coarser data sampling, signal-to-noise, and the use of Gaussian processes. Reproducing this large low-frequency lag with the disk+BLR model requires a slightly larger BLR radius than that inferred from the observed H$\beta$. We propose that the soft X-rays lagging the UV could be due to light travel time delays between the hard X-ray corona and distant photoionized gas that dominates the soft X-ray spectrum below 2~keV. 

\end{enumerate}
\begin{acknowledgments}
EK acknowledges support from NASA grant 80NSSC20K0802. EK, JG, CP acknowledge support through NASA grant 80NSSC22K1120. EMC gratefully acknowledges support from the NSF through grant No. AST-1909199. This work makes use of observations from the Las Cumbres Observatory global telescope network. We thank the anonymous referee for their meaningful contribution to the paper. We thank Aaron Barth and John Montano for their laudable work on the intercalibration of the ground-based telescope data used in the paper.
\end{acknowledgments}
\newpage

\bibliography{reference}{}
\bibliographystyle{aasjournal}

\end{document}